# Spacetime structure and asymmetric metric from the premetric formulation of electromagnetism

Wei-Tou Ni

*Center for Gravitation and Cosmology, Department of Physics,
National Tsing Hua University, Hsinchu, Taiwan, 30013, Republic of China*

**Abstract**

After reviewing the construction of metric from premetric electrodynamics with empirical improvement of accuracy in the skewonless case, we explore the role of skewons in the construction of spacetime metric in the full premetric electrodynamics. For metric principal part plus skewon part, we have shown that the Type I skewon part is constrained to $<$ a few $\times\ 10^{-35}$ in the weak field/weak EEP violation limit. Type II skewon part is not constrained in the first order. However, in the second order it induces birefringence; the nonbirefringence observations constrain the Type II skewon part to $\sim 10^{-19}$. Nevertheless, an additional nonmetric induced second-order contribution to the principal part constitutive tensor compensates the Type II skewon birefringence and makes it nonbirefringent. This second-order contribution is just the extra piece to the core-metric principal constitutive tensor induced by the antisymmetric tensor. From the Tamm-Rubilar tensor density and the dispersion relation for the spacetime medium with the asymmetric-metric-induced constitutive tensor, we know that it is nonbirefringent. The antisymmetric metric induced constitutive tensor has a pseudoscalar part in the decomposition. The variation of his part is constrained by observation on cosmic polarization rotation to $<$ 0.03, and gives one constraint on the 6-degree-of-freedom antisymmetric metric. Further studies of these 6 degrees of freedom experimental and theoretically are desired.

# 1. Introduction

## 1.1. Premetric formulation of electromagnetism

In the historical development, special relativity arose from the invariance of Maxwell equations under Lorentz transformation. In 1908, Minkowski [1] further put it into 4-dimensional geometric form with a metric invariant under Lorentz transformation. The use of metric as dynamical gravitational potential [2] and the employment of Einstein Equivalence Principle for coupling gravity to matter [3] are two important cornerstones to build general relativity [2,4]. In putting Maxwell equations into a form compatible with general relativity, Einstein noticed that the equations can be formulated in a form independent of the metric gravitational potential in 1916 [5,6]. Weyl [7], Murnaghan [8], Kottler [9] and Cartan [10] further developed and clarified this resourceful approach.

Maxwell equations for macroscopic/spacetime electrodynamics in terms of independently measurable field strength $F_{kl}$ (*E*, *B*) and excitation (density with weight +1) $H^{ij}$ (*D*, *H*) do not need metric as primitive concept (See, e. g., Hehl and Obukhov [11]):

$$H^{ij}{}_{,j} = -4\pi J^i, \tag{1a}$$
$$e^{ijkl} F_{jk,l} = 0, \tag{1b}$$

with $J^k$ the charge 4-current density and $e^{ijkl}$ the completely anti-symmetric tensor density of weight +1 ($e^{0123} = 1$). We use units with the light velocity *c* equal to 1. To complete this set of equations, a constitutive relation is needed between the excitation and the field:

$$H^{ij} = (1/2) \chi^{ijkl} F_{kl}. \tag{2}$$

Both $H^{ij}$ and $F_{kl}$ are antisymmetric, hence $\chi^{ijkl}$ must be antisymmetric in *i* and *j*, and in *k* and *l*. Hence the constitutive tensor density $\chi^{ijkl}$ (with weight +1) has 36 independent components, and can be uniquely decomposed into principal part (P), skewon part (Sk) and axion part (A) as given in [11, 12]:

$$\chi^{ijkl} = {}^{(P)}\chi^{ijkl} + {}^{(Sk)}\chi^{ijkl} + {}^{(Ax)}\chi^{ijkl}, \qquad (\chi^{ijkl} = -\chi^{jikl} = -\chi^{ijlk}) \tag{3}$$

with

$$^{(P)}\chi^{ijkl} = (1/6)[2(\chi^{ijkl} + \chi^{klij}) - (\chi^{iklj} + \chi^{ljik}) - (\chi^{iljk} + \chi^{jkil})], \tag{4a}$$

$$^{(Ax)}\chi^{ijkl} = \chi^{[ijkl]} = \varphi\, e^{ijkl}, \tag{4b}$$

$$^{(Sk)}\chi^{ijkl} = (1/2)\,(\chi^{ijkl} - \chi^{klij}). \tag{4c}$$

The principal part has 20 degrees of freedom. The axion part has one degree of freedom. The Hehl-Obukhov-Rubilar skewon part (4c) can be represented as

$$^{(Sk)}\chi^{ijkl} = e^{ijmk} S_m{}^l - e^{ijml} S_m{}^k, \tag{5}$$

with $S_m{}^n$ a traceless tensor of 15 independent degrees of freedom [11].

There are two equivalent definitions of constitutive tensor which are useful in various discussions (See, e. g., Hehl and Obukhov [11]). The first one is to take a dual on the first 2 indices of $\chi^{ijkl}$:

$$\kappa_{ij}{}^{kl} \equiv (1/2)\underline{e}_{ijmn}\,\chi^{mnkl}, \tag{6}$$

where $\underline{e}_{ijmn}$ is the completely antisymmetric tensor density of weight $-1$ with $\underline{e}_{0123} = 1$. Since $e_{ijmn}$ is a tensor density of weight $-1$ and $\chi^{mnkl}$ a tensor density of weight $+1$, $\kappa_{ij}{}^{kl}$ is a tensor. From (6), we have

$$\chi^{mnkl} = (1/2)\underline{e}^{ijmn}\kappa_{ij}{}^{kl}. \tag{7}$$

With this definition of constitutive tensor $\kappa_{ij}{}^{kl}$, the constitutive relation (2) becomes

$$*H_{ij} = (1/2)\,\kappa_{ij}{}^{kl}\,F_{kl}, \tag{8}$$

where $*H_{ij}$ is the Hodge dual of $H^{ij}$, i.e.

$$*H_{ij} \equiv (1/2)\,\underline{e}_{ijmn}\,H^{mn}. \tag{9}$$

The second equivalent definition of the constitutive tensor is to use a $6\times6$ matrix representation $\kappa_I{}^J$. Since $\kappa_{ij}{}^{kl}$ is nonzero only when the antisymmetric pairs of indices ($ij$) and ($kl$) have values (01), (02), (03), (23), (31), (12), the index pairs can be enumerated by capital letters $I$, $J$, … from 1 to 6 to obtain $\kappa_I{}^J$ ($\equiv \kappa_{ij}{}^{kl}$). With the relabeling, $F_{ij} \to F_I$, $H^{ij} \to H^I$, $\underline{e}_{ijmn} \to \underline{e}_{IJ}$, $e^{ijmn} \to e^{IJ}$. We have $F_I = (\boldsymbol{E}, -\boldsymbol{B})$ and $*H^I = (-\boldsymbol{H}, \boldsymbol{D})$. $\underline{e}_{IJ}$ and $e^{IJ}$ can be expressed in matrix form as

$$\underline{e}_{IJ} = e^{IJ} = \begin{bmatrix} 0 & \mathbf{I}_3 \\ \mathbf{I}_3 & 0 \end{bmatrix}, \tag{10}$$

where $\mathbf{I}_3$ is the 3 × 3 unit matrix. In terms of this definition, the constitutive relation (8) becomes

$$*H_I = \kappa_I{}^J F_J, \tag{11}$$

where $*H_I \equiv *H_{ij} = e_{IJ} H^J$. The axion part $^{(\text{Ax})}\chi^{ijkl}$ ($= \varphi\, e^{ijkl}$) now corresponds to

$$^{(\text{Ax})}\kappa_I{}^J = \varphi \begin{bmatrix} \mathbf{I}_3 & 0 \\ 0 & \mathbf{I}_3 \end{bmatrix} = \varphi\, \mathbf{I}_6, \tag{12}$$

where $\mathbf{I}_6$ is the 6 × 6 unit matrix. The principal part and the axion part of the constitutive tensor all satisfy the following equation (the skewonless condition):

$$e^{KJ}\kappa_J{}^I = \underline{e}^{IJ} \kappa_J{}^K. \tag{13}$$

*1.2. Construction of spacetime structure from premetric electrodynamics in the skewonless case*

In this subsection, we review the construction of spacetime structure from premetric electrodynamics. The first issue here is that how to (with what conditions can we) reach a metric or, owing to conformal invariance, how to reach a Riemannian light cone (a core metric up to conformal invariance) from the constitutive tensor. This issue has been studied rather thoroughly in the skewonless case, i.e. in the case that $\chi^{ijkl}$ is symmetric under the exchange of the index pairs $ij$ and $kl$ (in terms of the 6×6 matrix representation $\kappa_I{}^J$, equation (13) is satisfied) and has 20 principal components and 1 axionic component. In this case, the Maxwell equations can be derived from the Lagrangian $L$ ($= L_{\text{I}}^{(\text{EM})} + L_{\text{I}}^{(\text{EM-P})}$) with the electromagnetic field Lagrangian $L_{\text{I}}^{(\text{EM})}$ and the field-current interaction Lagrangian $L_{\text{I}}^{(\text{EM-P})}$ given by

$$L_{\text{I}}^{(\text{EM})} = -(1/(8\pi))H^{ij} F_{ij} = -(1/(16\pi))\chi^{ijkl} F_{ij} F_{kl}, \tag{14}$$
$$L_{\text{I}}^{(\text{EM-P})} = -A_k J^k, \tag{15}$$

with $\chi^{ijkl} = -\chi^{jikl}$ a tensor density of the gravitational fields or matter fields to be investigated, $F_{ij} \equiv A_{j,i} - A_{i,j}$ the electromagnetic field strength tensor, $A_i$ the electromagnetic 4-potential guaranteed by the second Maxwell equation (1b), and comma denoting partial derivation. We note that only the part of $\chi^{ijkl}$ which is symmetric under the interchange of index pairs *ij* and *kl* contributes to the Lagrangian, i.e. skewon part does not contribute and we assume it is absent in this subsection. Three conditions have been studied for this symmetric constitutive tensor:

(i) The closure condition: Toupin [13], Schönberg [14], and Jadczyk [15] have investigated this approach. The closure condition on the skewonless constitutive tensor is

$$\kappa\,\kappa = (\kappa_I{}^J \kappa_J{}^K) = (1/6)\,\mathrm{tr}(\kappa\,\kappa)\,\mathbf{I}_6. \qquad (16)$$

With this closure condition, it has been shown that the constitutive tensor must be metric with a dilatonic degree of freedom $\psi$, i.e.,

$$\chi^{ijkl} = (-h)^{1/2}[(1/2)h^{ik} h^{jl} - (1/2)h^{il} h^{kj}]\psi, \qquad (17)$$

where $h^{ij}$ is a symmetric metric with inverse $h_{ij}$, $h = \det(h_{ij})$, and $\psi$ is a scalar (dilaton) degree of freedom.

(ii) The Galileo weak equivalence principle: In the 1970s, we used Galileo Equivalence Principle and derived its consequences for an electromagnetic system with the particle Lagangian $L_I^{(P)}$ [16,17]:

$$L_I^{(P)} = -\Sigma_I m_I (ds_I)/(dt)\,\delta(\boldsymbol{x}-\boldsymbol{x}_I). \qquad (18)$$

Here $m_I$ is the mass of the *I*th (charged) particle and $s_I$ its 4-line element from the metric $g_{ij}$. The result is that the constitutive tensor density $\chi^{ijkl}$ and the particle metric must satisfy the following relation:

$$\chi^{ijkl} = (-g)^{1/2}[(1/2)\,g^{ik} g^{jl} - (1/2)\,g^{il} g^{kj}] + \varphi\,e^{ijkl}, \qquad (19)$$

where $g^{ij}$ is the inverse of the particle metric $g_{ij}$, $g = \det(g_{ij})$, and $\varphi$ is a pseudoscalar (axion) degree of freedom in the relation. Since it is well-known that the axion degree of freedom does not affect the propagation of light in the lowest eikonal approximation [11,18-24], the particle metric $g^{ij}$ (or $g_{ij}$) also generates the light cone for electromagnetic wave propagation in this approximation.

(iii) The nonbirefringence condition: The third approach used empirical observations/experiments to constrain the constitutive tensor density interaction [18-20]. From the nonbirefringence observations of electromagnetic wave propagation in spacetime, we constructed the light cone and constrained the constitutive tensor to metric form compatible with this light cone plus an additional scalar (dilaton) field and an additional pseudoscalar (axion) field to high precision [18-20]. The theoretical condition for no birefringence (no splitting, no retardation) for electromagnetic wave propagation in all directions is that the constitutive tensor density $\chi^{ijkl}$ can be written in the following form

$$\chi^{ijkl} = (-h)^{1/2}[(1/2)\, h^{ik}\, h^{jl} - (1/2)\, h^{il}\, h^{kj}]\psi + \varphi\, e^{ijkl}, \qquad (20)$$

where $h^{ij}$ is a metric constructed from $\chi^{ijkl}$ ($h$ = det ($h_{ij}$) and $h_{ij}$ the inverse of $h^{ij}$) which generates the light cone for electromagnetic wave propagation [18-21].

We constructed the relation (20) in the weak-violation approximation of the Einstein Equivalence Principle (EEP) in 1981 [18-20]; Haugan and Kauffmann [21] reconstructed the relation (20) in 1995. After the cornerstone work of Lämmerzahl and Hehl [22], Favata [25] finally proves the relation (20) without assuming weak-field approximation (see also Dahl [26]). Polarization measurements of electromagnetic waves from pulsars and cosmologically distant astrophysical sources has yielded stringent constraints agreeing with (20) down to $10^{-16}$ and $10^{-32}$ respectively (for a review, see [24]). Recent polarization observations on gamma-ray bursts gives even better constraints on nonbirefringence in cosmic propagation [27,28]. The observation on the polarized gamma-ray burst GRB 061122 ($z$ = 1.33) gives a lower limit on its polarization fraction of 60% at 68% confidence level (c.l.) and 33% at 90% c.l. in the 250-800 keV energy range [27]. The observation on the polarized gamma-ray burst GRB 140206A constrains the linear polarization level of the second peak of this GRB above 28 % at 90% c.l. in the 200-400 keV energy range [28]; the redshift of the source is measured from the GRB afterglow optical spectroscopy to be $z$ = 2.739. Since birefringence is proportional to the wave vector $k$ in our case, as gamma-ray of a particular frequency (energy) travels in the cosmic spacetime, the two linear polarization eigen-modes would pick up small phase differences. A linear polarization mode from distant source resolved into these two modes will become elliptical polarized during travel and lose part of the linear coherence. The way of gamma ray losing linear coherence depends on the frequency span. For a band of frequency, the extent of losing coherence depends on the distance of travel. The depolarization distance is of the order of frequency band span $\pi\Delta f$ times the integral $I = \int(1 + z(t))dt$ of the redshift factor $(1 + z(t))$ with respect to the time of travel. For GRB 140206A, this is about

$$\pi\Delta f\, I = \pi\Delta f \int(1 + z(t))dt \approx 1.5 \times 10^{20}\ \text{Hz} \times 0.6 \times 10^{18}\ \text{s} \approx 10^{38}. \qquad (21)$$

Since we do observe linear polarization in the 200-400 kHz frequency band of GRB 140206A with lower bound of 28 %, this gives a fractional constraint of about $10^{-38}$ on a combination of $\chi$'s. A similar constraint can be obtained for GRB 061122 (the band width times the redshift is about the same). A more detailed modeling may give better limits. The distribution of GRBs is basically isotropic. When this procedure is applied to an ensemble of polarized GRBs from various directions, the relation (20) would be verified to about $10^{-38}$. For a more detailed discussion, please see [29].

Nonbirefringence (no splitting, no retardation) for electromagnetic wave propagation independent of polarization and frequency (energy) in all directions can be formulated as a statement of Galilio Equivalence Principle for photons. However, the complete agreement with EEP for photon sector requires (i) no birefringence; (ii) no polarization rotation; (iii) no amplification/no attenuation in spacetime propagation. With nonbirefringence, any skewonless spacetime constitutive tensor must be of the form (20) as we just reviewed. Assuming (20) in [30], we have shown that with the condition of no amplification/no attenuation and the condition of no polarization rotation satisfied, the the dilaton $\psi$ and axion $\varphi$ should be constant respectively. That is, no varying dilaton field and no varying axion field respectively; the EEP for photon sector is observed; the spacetime constitutive tensor is of metric-induced form. Thus we tie the three observational conditions to EEP and to metric-induced spacetime constitutive tensor in the photon sector. We summarize the accuracies of verifying these three empirical conditions in Table I.

Table I. Constraints on the spacetime constitutive tensor $\chi^{ijkl}$ and construction of the spacetime structure (metric + axion field $\varphi$ + dilaton field $\psi$) from experiments/observations in skewonless case ($U$: Newtonian gravitational potential)

| Experiment | Constraints | Accuracy |
|---|---|---|
| Pulsar Signal Propagation | | $10^{-16}$ |
| Radio Galaxy Observation | $\chi^{ijkl} \rightarrow \tfrac{1}{2}\,(-h)^{1/2}[h^{ik}\,h^{jl} - h^{il}\,h^{kj}]\psi + \varphi e^{ijkl}$ | $10^{-32}$ |
| Gamma Ray Burst (GRB) | | $10^{-38}$ |
| CMB Spectrum Measurement | $\psi \rightarrow 1$ | $8 \times 10^{-4}$ |
| Cosmic Polarization Rotation Experiment | $\varphi - \varphi_0\,(\equiv \alpha) \rightarrow 0$ | $|<\alpha>| < 0.02$, $<(\alpha - <\alpha>)^2>^{1/2} < 0.03$ |
| Eötvös-Dicke-Braginsky Experiments | $\psi \rightarrow 1$ <br> $h_{00} \rightarrow g_{00}$ | $10^{-10}\,U$ <br> $10^{-6}\,U$ |
| Vessot-Levine Redshift Experiment | $h_{00} \rightarrow g_{00}$ | $1.4 \times 10^{-4}\,U$ |
| Hughes-Drever-type Experiments | $h_{ij} \rightarrow g_{ij}$ <br> $h_{0i} \rightarrow g_{0i}$ <br> $h_{00} \rightarrow g_{00}$ | $10^{-18}\,U$ <br> $10^{-13}$-$10^{-14}\,U$ <br> $10^{-10}\,U$ |

The birefringence condition is the condition amenable to the most precise observations. The constraint listed in the third row on the dilaton field is from the agreement of the cosmic microwave background (CMB) to blackbody radiation and its precise calibration [30]. The constraints listed on the axion field are from the UV polarization observations of radio galaxies and the CMB polarization observations [31,32,23,24] -- 0.02 for Cosmic Polarization Rotation (CPR) mean value $|<\alpha>|$ and 0.03 for the CPR fluctuations $<(\alpha - <\alpha>)^2>^{1/2}$.

We further constrained the light cone metric to matter metric up to a scalar factor from Hughes-Drever-type experiments and the dilaton to 1 (constant) from Eötvös-type experiments to high precision [18,19,24,33]. These analyses apply to weak field and strong field equally well. In summary, we can construct the spacetime metric plus mild constraints on axion (pseudoscalar) degree of freedom from experiments/observations in the constitutive tensor framework in the skewonless case as in Table I. The constraint on dilaton before the last scattering surface of CMB needs more study in view of various dilatonic inflationary theories.

In section 2, we discuss the empirical foundation of the closure relation. In section 3, we study the constraints on Type II skewon field to second order. Although the constitutive tensor with the metric principal part and Type II skewon part is birefringent in the second order, an added 2$^{nd}$ order nonmetric piece to the metric principal part would make the constitutive tensor nonbirefringent. This added part turns out to be just the extra piece in the principal part of an asymmetric-metric induced constitutive tensor. In section 4, we turn to the study of dispersion relation of an asymmetric-metric induced constitutive tensor. In section 5, we study the experimental/observational constraints on the asymmetric-metric induced spacetime constitutive tensor. In section 6, we present an outlook and a few discussions.

**2. Empirical foundation of the closure relation for skewonless case**

In terms of $\kappa_{ij}{}^{kl}$ (defined in (6)) and re-indexed $\kappa_I{}^J$, the constitutive tensor (20) is represented in the following forms:

$$\kappa_{ij}{}^{kl} = (1/2)\, \underline{e}_{ijmn}\, \chi^{mnkl} = (1/2)\, \underline{e}_{ijmn}\, (-h)^{1/2}\, h^{mk}\, h^{nl}\, \psi + \varphi\, \delta_{ij}{}^{kl}, \tag{22}$$

$$\kappa_I{}^J = (1/2)\, \underline{e}_{ijmn}\, (-h)^{1/2}\, h^{mk}\, h^{nl}\, \psi + \varphi\, \delta_I{}^J, \tag{23}$$

where $\delta_{ij}{}^{kl}$ is a generalized Kronecker delta defined as

$$\delta_{ij}{}^{kl} = \delta_i{}^k\, \delta_j{}^l - \delta_i{}^l\, \delta_j{}^k. \tag{24}$$

In the derivation, we have used the formula

$$\underline{e}_{ijmn} \, e^{mnkl} = 2 \, \delta_{ij}{}^{kl}. \tag{25}$$

Let us calculate $\kappa_{ij}{}^{kl}\kappa_{kl}{}^{pq}$ for the constitutive tensor (22):

$$\begin{aligned}\kappa_{ij}{}^{kl} \, \kappa_{kl}{}^{pq} &= [(1/2) \, \underline{e}_{ijmn} \, (-h)^{1/2} \, h^{mk} \, h^{nl} \, \psi + \varphi \, \delta_{ij}{}^{kl}] \, [(1/2) \, \underline{e}_{klrs} \, (-h)^{1/2} \, h^{rp} \, h^{sq} \, \psi + \varphi \, \delta_{kl}{}^{pq}] \\ &= -(1/2) \, \delta_{ij}{}^{pq}\psi^2 + 2 \, \delta_{ij}{}^{pq}\varphi^2 + \underline{e}_{ijrs} \, (-h)^{1/2} \, h^{rp} \, h^{sq} \, \varphi \, \psi \\ &= -(1/2) \, \delta_{ij}{}^{pq}\psi^2 + 2 \, \varphi \, {}^{(P)}\kappa_{ij}{}^{pq} \end{aligned} \tag{26}$$

where we have used (25) and the following relations

$$e_{klrs} \, h^{mk} \, h^{nl} \, h^{rp} \, h^{sq} = e^{mnpq} \det(h^{uv}), \tag{27}$$
$$\det(h^{uv}) = [\det(h_{uv})]^{-1} = h^{-1}, \tag{28}$$
$$\delta_{ij}{}^{kl} \, \delta_{kl}{}^{pq} = 2 \, \delta_{ij}{}^{pq}. \tag{29}$$

In terms of the six-dimensional index $I$, equation (26) becomes

$$\kappa_I{}^J \, \kappa_J{}^K = (1/2) \, \kappa_{ij}{}^{kl} \, \kappa_{kl}{}^{pq} = \delta_I{}^K \, [(1/4)\psi^2] + {}^{(P)}\kappa_I{}^K \, \varphi = (1/4) \, \psi^2 \, \delta_I{}^J + \kappa_I{}^K \, \varphi. \tag{30}$$

Thus the matrix multiplication of $\kappa_I{}^J$ with itself is a linear combination of itself and the identity matrix, and generates a closed algebra of linear dimension 2. The algebraic relation (30) is a closure relation that generalizes the closure relation (16). The matrix multiplication of $\kappa_I{}^J$ satisfies the closure relation (30). In case $\varphi = 0$, the axion part ${}^{(Ax)}\kappa_I{}^J$ of the constitutive tensor vanishes and (30) reduces to closure relation (16).

From the nonbirefringence condition (20), we derive the closure relation (30) in a number of algebraic steps which consist of order 100 individual operations of addition/subtraction or multiplication. Equation (20) is empirically verified to $10^{-38}$. Therefore equation (17) is empirically verified to $10^{-37}$ (precision $10^{-38}$ times $100^{1/2}$). Therefore, when there are no axion and no dilaton, the closer relation (16) is empirically verified to $10^{-37}$. Since dilaton is constrained to $8 \times 10^{-4}$, if one allow for dilaton, relation (9) is verified to $8 \times 10^{-4}$ since the last scattering surface of CMB; since axion is constrained to $10^{-2}$, if one allow for axion in addition, relation (9) is verified to $10^{-2}$ since the last scattering surface of CMB.

## 3. Constraints on skewon field to second order

The skewon field can be decomposed into two types – Type I and Type II – by using the Minkowski metric $\eta^{ij}$ (or metric $g^{ij}$ or $h^{ij}$) to raise and lower indices of $S_m{}^n$:

I. Type I Skewon field: $^{(SkI)}\chi^{ijkl}$ with symmetric $^{(SkI)}S_{mn}$;
II. Type II Skewon field: $^{(SkII)}\chi^{ijkl}$ with anti-symmetric $^{(SkII)}S_{mn}$,

where

$$S_{mn} \equiv S_m{}^i \eta_{in} \text{ (or } g_{ij} \text{ or } h^{ij}); \, ^{(SkI)}S_{mn} \equiv (1/2)(S_{mn} + S_{nm}); \, ^{(SkII)}S_{mn} \equiv (1/2)(S_{mn} - S_{nm}), \quad (31a)$$

$$S_{mn} = {}^{(SkI)}S_{mn} + {}^{(SkII)}S_{mn}. \quad (31b)$$

This classification is invariant under tensor transformation. General skewon field $^{(Sk)}\chi^{ijkl}$ can be written as the sum of two parts, i.e., $^{(SkI)}\chi^{ijkl} + {}^{(SkII)}\chi^{ijkl}$. Type I skewon field has 9 degrees of freedom. Type II skewon field has 6 degrees of freedom. The astrophysical and cosmological constraints on skewon field in spacetime electrodynamics have been discussed and studied in [34, 35]. We will discuss this issue in detail and put further constraints on it.

### *3.1. Constraints on skewon field to first order in the weak field limit*

In a previous paper [35], we extended the weak-field analysis [18-20] to include the skewons. From the dispersion relation we show that no dissipation/no amplification condition implies that the additional skewon field must be of Type II. For Type I skewon field, the dissipation/amplification is proportional to the frequency and the CMB spectrum would deviate from Planck spectrum. From the high precision agreement of the CMB spectrum to 2.755 K Planck spectrum, we constrain the Type I cosmic skewon field $|^{(SkI)}\chi^{ijkl}|$ to $\leq$ a few $\times$ $10^{-35}$. The skewon part of constitutive tensor constructed from asymmetric metric is of Type II, hence is allowed in this approximation.

The dispersion relation we have derived in [35] is

$$\omega = k[1 + (1/2)(A_{(1)} + A_{(2)}) \pm (1/2)((A_{(1)} - A_{(2)})^2 + 4B_{(1)}B_{(2)})^{1/2}] + O(2), \quad (32)$$

where

$$A_{(1)} \equiv \chi^{(1)1010} - (\chi^{(1)1013} + \chi^{(1)1310}) + \chi^{(1)1313}, \quad (33a)$$
$$A_{(2)} \equiv \chi^{(1)2020} - (\chi^{(1)2023} + \chi^{(1)2320}) + \chi^{(1)2323}, \quad (33b)$$
$$B_{(1)} \equiv \chi^{(1)1020} - (\chi^{(1)1023} + \chi^{(1)1320}) + \chi^{(1)1323}, \quad (33c)$$
$$B_{(2)} \equiv \chi^{(1)2010} - (\chi^{(1)2013} + \chi^{(1)2310}) + \chi^{(1)2313}. \quad (33d)$$

The quantity under the square root sign is

$$\xi \equiv (A_{(1)} - A_{(2)})^2 + 4B_{(1)} B_{(2)} = (A_{(1)} - A_{(2)})^2 + 4(^{(P)}B)^2 - 4(^{(Sk)}B)^2. \tag{34}$$

Depending on the sign or vanishing of $\xi$, we have the following three cases of electromagnetic wave propagation:

(i) $\xi > 0$, $(A_{(1)} - A_{(2)})^2 + 4(^{(P)}B)^2 > 4(^{(Sk)}B)^2$: There is birefringence of wave propagation;

(ii) $\xi = 0$, $(A_{(1)} - A_{(2)})^2 + 4(^{(P)}B)^2 = 4(^{(Sk)}B)^2$: There are no birefringence and no dissipation/amplification in wave propagation;

(iii) $\xi < 0$, $(A_{(1)} - A_{(2)})^2 + 4(^{(P)}B)^2 < 4(^{(Sk)}B)^2$: There is no birefringence, but there are both dissipative and amplifying modes in wave propagation.

When the principal part $^{(P)}\chi^{ijkl}$ of the constitutive tensor is given by (17), it is easy to check by substitution that

$$A_{(1)} = A_{(2)} \text{ and } {}^{(P)}B_{(1)} = 0. \tag{35}$$

In this case, (i) does not happen, (ii) and (iii) become

(ii)' $\xi = 0$, $^{(Sk)}B = 0$: There are no birefringence and no dissipation/amplification in wave propagation;

(iii)' $\xi < 0$, $^{(Sk)}B \neq 0$: There is no birefringence, but there are both dissipative and amplifying modes in wave propagation.

For type II skewon field, we have $^{(Sk)}B = 0$; therefore it is the case (ii)'. In Sec.IV.A.1 of Obukhov and Hehl [34], the type II skewon field with the core-metric principal part gives birefringence in the second order of the skewon field strength and there is no birefringence in the first order consistent with our first-order result.

*3.2. Constraints on skewon field to second order in the weak field limit*

The dispersion relation for the wave covector $q_i$ of electromagnetic propagation with general constitutive tensor (3) in the geometric-optics limit is given by *the generalized covariant Fresnel equation* [11]:

$$G^{ijkl}(\chi) q_i q_j q_k q_l = 0, \tag{36}$$

where $G^{ijkl}(\chi) (= G^{(ijkl)}(\chi))$ is a completely symmetric fourth order Tamm-Rubilar (TR) tensor density of weight +1 defined by

$$G^{ijkl}(\chi) \equiv (1/4!)\, \underline{e}_{mnpq}\, \underline{e}_{rstu}\, \chi^{mnr(i} \chi^{j|ps|k} \chi^{l)qtu}. \tag{37}$$

For the decomposition (3), Hehl, Obukhov and Rubilar [12] showed that

$$G^{ijkl}(\chi) = G^{ijkl}(^{(P)}\chi) + {}^{(P)}\chi^{mr(i|n|j}S_m{}^k S_n{}^l. \tag{38}$$

For constitutive tensor consists of metric-induced part (metric: $g^{ij}$) and Type II skewon with $S^{mk}$ ($\equiv g^{ml} S_l{}^k = a^{mk} = -a^{km}$), Obukhov and Hehl [34] showed that its Tamn-Rubilar tensor density is

$$G^{ijkl}(\chi) = -(-g)^{1/2}g^{(ij}g^{kl)} + {}^{(P)}\chi^{mr(i|n|j}a_m{}^k a_n{}^l. \tag{39}$$

To the first order of $a_m{}^k$ (or $S_m{}^k$), there is nonbirefringence. However, in the second order, Obukhov and Hehl [35] showed that the light propagation is birefringent with one light cone defined by the spacetime metric $g^{ij}$ and the second by the metric $^{(2)}g^{ij}$:

$$^{(2)}g^{ij} = g^{ij} - a^{mi}a_m{}^j. \tag{40}$$

The dispersion relations for the wave propagation in the spacetime metric and in the second optical metric are

$$g^{ij} q_i q_j = 0, \tag{41a}$$
$$^{(2)}g^{ij} q_i q_j = g^{ij} q_i q_j - a_{mi}a^{mj} q_i q_j = 0. \tag{41b}$$

Let us calculate the birefringence of this constitutive tensor for plane wave propagating in the z-direction. We use a local inertial frame for calculation. The wave vector $q^i$ and covector $q_i$ for this plane wave are

$$q^i = (\omega, 0, 0, k); q_i = (\omega, 0, 0, -k). \tag{42}$$

From equations (41a,b), we obtain

$$\omega_{(1)} = k, \tag{43a}$$
$$\omega_{(2)} = k \{1 - (1/2) [(a^{10} - a^{13})^2 + (a^{20} - a^{23})^2] + O(a^4)\}. \tag{43b}$$

Therefore the birefringence index is

$$\Delta n = n_{(1)} - n_{(2)} = \omega_{(1)}/k - \omega_{(2)}/k = (1/2) [(a^{10} - a^{13})^2 + (a^{20} - a^{23})^2] < 10^{-38}, \tag{44a}$$

and

$$|a^{10} - a^{13}| < 1.4 \times 10^{-19}; |a^{20} - a^{23}| < 1.4 \times 10^{-19}. \qquad (44b)$$

Hence we have

$$a^{10} = a^{13}; a^{20} = a^{23} \text{ to } 1.4 \times 10^{-19} \qquad (45).$$

With nonbirefringence in all directions, we can show that all $a^{ij}$ are of the order of a few $\times 10^{-19}$: (i) First, by a rotation in the yz-plane of 180° angle, we have $a^{10} = -a^{13}$ and $a^{20} = -a^{23}$; (ii) Second, by adding or subtracting (45), we have $a^{10} = a^{13} = a^{20} = a^{23} = 0$ to $10^{-19}$; (iii) Then, by permutation in space indices, we have $a^{30} = a^{21} = 0$ to $10^{-19}$. Therefore all $a^{ij} = 0$ to $10^{-19}$.

The birefringence (44a) is second order in $a_{mi}$ ($S_m^i$). From the observation on gamma-ray burst, Obukhov and Hehl have obtained "the estimates $s_a{}^b s_b{}^a \sim a_a{}^b a_b{}^a < 7 \times 10^{-27} \lambda_0^2$ and $n_a n^a < 3 \times 10^{-27} \varepsilon_0^2$" ($a, b, c,\ldots = 1, 2, 3$ are the 3 indices and $n_a = S_a^0$; in our units, $\lambda_0$ and $\varepsilon_0$ are 1) and conclude that, "if the skewon indeed spoils the light cone structure, its influence should be extremely small" in their pioneer work. From improved gamma ray observations (assuming that the observed direction is not special) and more detailed analysis, we have shown that the Type II skewons are constrained to $10^{-19}$ and the square combinations of $a$'s are constrained to $O(10^{-38})$ if the principal part is given by the metric core form (20). However, *if the principal part has some additional part to cancel the skewon contribution to birefringence/amplification-dissipation, would there be no birefringence/no amplification-dissipation?* This requires the effect of added principal part cancels the effect of skewon part in $\xi$ in the second order.

Before we turn into this issue, we summarize various 1st-order and 2nd-order effects in wave propagation on media with the core-metric based constitutive tensors in Table II.

## 4. Constitutive tensor from asymmetric metric and Fresnel equation

Eddington [36], Einstein & Straus [37], and Schrödinger [38,39] considered asymmetric metric in their exploration of gravity theories. Just like we can build spacetime constitutive tensor from the (symmetric) metric as in metric theories of gravity, we can also build it from the asymmetric metric. Let $q^{ij}$ be the asymmetric metric as follows:

$$\chi^{ijkl} = \tfrac{1}{2} \, (-q)^{1/2}(q^{ik}q^{jl} - q^{il}q^{jk}), \tag{46}$$

with $q = \det^{-1}({}^{(S)}q^{ij})$. When $q^{ij}$ is symmetric, this definition reduces to that of the metric theories of gravity. Resolving the asymmetric metric into symmetric part ${}^{(S)}q^{ij}$ and anti-symmetric part ${}^{(A)}q^{ij}$:

$$q^{ij} = {}^{(S)}q^{ij} + {}^{(A)}q^{ij}, \text{ with } {}^{(S)}q^{ij} \equiv \tfrac{1}{2}(q^{ij}+q^{ji}) \text{ and } {}^{(A)}q^{ij} \equiv \tfrac{1}{2}(q^{ij}-q^{ji}), \tag{47}$$

we can decompose the constitutive tensor into the principal part ${}^{(P)}\chi^{ijkl}$, the axion part ${}^{(Ax)}\chi^{ijkl}$ and skewon part ${}^{(Sk)}\chi^{ijkl}$ as follows [35,40]:

$$\chi^{ijkl} = \tfrac{1}{2} \, (-q)^{1/2}(q^{ik}q^{jl}-q^{il}q^{jk}) = {}^{(P)}\chi^{ijkl} + {}^{(Ax)}\chi^{ijkl} + {}^{(Sk)}\chi^{ijkl}, \tag{48a}$$

with

$${}^{(P)}\chi^{ijkl} \equiv \tfrac{1}{2}(-q)^{1/2}\,({}^{(S)}q^{ik}\,{}^{(S)}q^{jl} - {}^{(S)}q^{il}\,{}^{(S)}q^{jk} + {}^{(A)}q^{ik}\,{}^{(A)}q^{jl} - {}^{(A)}q^{il}\,{}^{(A)}q^{jk} - 2\,{}^{(A)}q^{[ik}\,{}^{(A)}q^{jl]}), \tag{48b}$$

$${}^{(Ax)}\chi^{ijkl} \equiv (-q)^{1/2}\,{}^{(A)}q^{[ik}\,{}^{(A)}q^{jl]}, \tag{48c}$$

$${}^{(Sk)}\chi^{ijkl} \equiv \tfrac{1}{2}(-q)^{1/2}\,({}^{(A)}q^{ik}\,{}^{(S)}q^{jl} - {}^{(A)}q^{il}\,{}^{(S)}q^{jk} + {}^{(S)}q^{ik}\,{}^{(A)}q^{jl} - {}^{(S)}q^{il}\,{}^{(A)}q^{jk}). \tag{48d}$$

The axion part ${}^{(Ax)}\chi^{ijkl}$ only comes from the second order terms of ${}^{(A)}q^{il}$.

Table II. Various 1st-order and 2nd-order effects in wave propagation on media with the core-metric based constitutive tensors. ${}^{(P)}\chi^{(c)}$ is the extra piece [last 3 terms in (48b)] to the core-metric principal part for canceling the skewon contribution to birefringence/amplification-dissipation.

| Constitutive tensor | Birefringence (in the geometric optics approximation) | Dissipation/ amplification | Spectroscopic distortion | Cosmic polarization rotation |
|---|---|---|---|---|
| Metric: $\tfrac{1}{2}(-h)^{1/2}[h^{ik}h^{jl} - h^{il}h^{kj}]$ | No | No | No | No |
| Metric + dilaton: $\tfrac{1}{2}(-h)^{1/2}[h^{ik}h^{jl} - h^{il}h^{kj}]\psi$ | No (to all orders in the field) | Yes (due to dilaton gradient) | No | No |
| Metric + axion: $\tfrac{1}{2}(-h)^{1/2}[h^{ik}h^{jl} - h^{il}h^{kj}] + \varphi e^{ijkl}$ | No (to all orders in the field) | No | No | Yes (due to axion gradient) |
| Metric + dilaton + axion: $\tfrac{1}{2}(-h)^{1/2}[h^{ik}h^{jl} - h^{il}h^{kj}]\psi + \varphi e^{ijkl}$ | No (to all orders in the field) | Yes (due to dilaton gradient) | No | Yes (due to axion gradient) |
| Metric + type I skewon | No to first order | Yes | Yes | No |
| Metric + type II skewon | No to first order; yes to 2nd order | No to first order and to 2nd order | No | No |
| Metric + ${}^{(P)}\chi^{(c)}$+ type II skewon | No to first order; no to 2nd order | No to first order and to 2nd order | No | No |
| Asymmetric metric induced: $\tfrac{1}{2}(-q)^{1/2}(q^{ik}q^{jl} - q^{il}q^{jk})$ | No (to all orders in the field) | No | No | Yes (due to axion gradient) |

Using $^{(S)}q^{ij}$ to raise and its inverse to lower the indices, we have as equation (16) in [35]

$$S_{ij} = \tfrac{1}{4}\ \varepsilon_{ijmk}\ ^{(A)}q^{mk};\ ^{(A)}q^{mk} = -\ \varepsilon^{mkij}\ S_{ij}, \tag{49}$$

where $\varepsilon_{ijmk}$ and $\varepsilon^{mkij}$ are respectively the completely antisymmetric covariant and contravariant tensors with $\varepsilon^{0123} = 1$ and $\varepsilon_{0123} = -1$ in local inertial frame. Thus the skewon field from asymmetric metric is of type II.

*Dispersion relation in the geometrical optics limit.* There are two ways to obtain the Tamm-Rubilar tensor density (37) for the dispersion relation (36). One way is by straightforward calculation; the other is by covariant method [40]. In the Appendix we outline the straightforward calculation to obtain the Tamm-Rubilar tensor density $G^{ijkl}(\chi)$ for the asymmetric metric induced constitutive tensor:

$$G^{ijkl}(\chi) = (1/8)\ (-q)^{3/2}\ \det(q^{ij})\ q^{(ij}q^{kl)} = (1/8)\ (-q)^{3/2}\ \det(q^{ij})\ ^{(S)}q^{(ij\ (S)}q^{kl)}. \tag{50}$$

Except for a scalar factor, (50) is the same as for metric-induced constitutive tensor with $^{(S)}q_{ij}$ replacing the metric $g_{ij}$ or $h_{ij}$. Therefore in the geometric optical approximation, there is no birefringence and the unique light cone is given by the metric $^{(S)}q_{ij}$.

## 5. Constraints on asymmetric-metric induced constitutive tensor

Although the asymmetric-metric induced constitutive tensor leads to a Fresnel equation which is nonbirefringent, it contains an axionic part:

$$^{(Ax)}\chi^{ijkl} \equiv (-q)^{1/2}\ ^{(A)}q^{[ik\ (A)}q^{jl]} = \varphi\ e^{ijkl};\ \varphi \equiv (1/4!)\ e_{ijkl}\ (-q)^{1/2}\ ^{(A)}q^{[ik\ (A)}q^{jl]}, \tag{54}$$

which induces polarization rotation in wave propagation. Constraints on cosmic polarization rotation (CPR) and its fluctuation constrain the axionic part and therefore also constrain the asymmetric metric. The variation of $\varphi$ ($\equiv (1/4!)\ e_{ijkl}\ (-q)^{1/2}\ ^{(A)}q^{[ik\ (A)}q^{jl]}$) is constrained by observations [31,32,23,24] on the cosmic polarization rotation to $< 0.02$ and its fluctuation to $< 0.03$ since the last scattering surface, and in turn constrains the antisymmetric metric of the spacetime for this degree of freedom. The antisymmetric metric has 6 degrees of freedom. Further study of the remaining 5 degrees of freedom experimentally and theoretically would be desired.

## 6. Discussion and outlook

In the following, we highlight a few points and discuss various issues to be studied further.

(i) The complete agreement with EEP for photon sector requires (i) no birefringence; (ii) no polarization rotation; (iii) no amplification/no attenuation in spacetime propagation. We summarize the empirical verifications of these conditions in Table I. Nonbirefringence (no splitting, no retardation) for electromagnetic wave propagation in all directions independent of polarization and frequency (energy) (Galilio Equivalence Principle for photons) is the basic condition. With no birefringence, any skewonless spacetime constitutive tensor must be of the core metric form (20) with a dilatonic freedom and an axionic freedom. With the recent polarization observation on GRBs, we have shown in Section 1.2 that the nonbirefringence condition is verified to $10^{-38}$ in the direction of observation. More of this kind of observations would verify the complete core metric form to $10^{-38}$ or better. From the core metric form, we have shown that the generalized closure relation (30) holds and is verified empirically to $10^{-37}$. Therefore with no (varying) dilaton and no (varying) axion, the original closure relation (16) is verified empirically to $10^{-37}$. This is probably the best empirical evidence for the closure relation. Further experiments/observations to look for birefringence, dilaton and axion are desired.

(ii) When skewon is an allowed piece in the constitutive tensor with the core-metric principal part, various effects in wave propagation for various combinatorial constitutive tensors are tabulated in Table II. Type I skewon field induces amplification/dissipation with frequency dependence/distortion in wave propagation, and is constrained to a few $\times 10^{-35}$ [35]. Type II skewon part with metric constitutive tensor is not constrained in the first order. In the second order it induces birefringence and is constrained to $< 10^{-19}$ from cosmic nonbirefringence observations. An additional nonmetric principal-part constitutive tensor from the extra piece induced by the antisymmetric metric tensor compensates the Type II skewon nonbirefringence and makes the whole constitutive tensor nonbirefringent.

(iii) The medium/spacetime with asymmetric-metric-induced constitutive tensor is nonbirefringent. However, its constitutive tensor has a pseudoscalar part in the decomposition. The time variation of this part is constrained by observation on the cosmic polarization rotation to $< 0.02$ since the last scattering surface while its space fluctuations are constrained to $< 0.03$ on the last scattering surface. In turn this degree of freedom of the antisymmetric metric of the spacetime is constrained correspondingly. Further studies on the constraints or evidence for this degree of freedom and the remaining 5 degrees of freedom are warranted.

(iv) Ever since the skewon field was proposed by Hehl, Obukhov and Rubilar [12, 42] and it has been studied extensively [43-48,35]. As demonstrated in these studies, skewons are rich in their properties. Further studies on general properties and specific examples may lead to better understanding and potential applications.

(v) In previous papers [49,50], we have explored the foundations of classical electrodynamics using $\chi$-framework. The results of this paper further support the empirical foundations of Maxwell-Lorentz theory of electromagnetism, especially the high accuracy verification of the generalized closure relation which is related to (generalized) electromagnetic duality.

**Acknowledgements**

We would like to thank the National Science Council (Grant No. NSC102-2112-M-007-019) and the National Center for Theoretical Sciences (NCTS) for supporting this work in part.

**Appendix. Calculation of the Tamm-Rubilar (TR) tensor density for the constitutive tensor induced by asymmetric metric**

In this Appendix, we calculate the Tamm-Rubilar (TR) tensor density for the constitutive tensor induced by asymmetric metric (48a-d). There are two ways to obtain the Tamm-Rubilar tensor density, by straightforward calculation or by covariant method. Here we calculate it straightforwardly. Define

$$
\begin{aligned}
Z^{ijkl}(\chi) &\equiv (1/4!)\, \underline{e}_{mnpq}\, \underline{e}_{rstu}\, \chi^{mnri} \chi^{jpsk} \chi^{lqtu}. \\
&= (1/4!)\,(1/8)(-q)^{3/2}\, \underline{e}_{mnpq}\, \underline{e}_{rstu}\, (q^{mr}q^{ni} - q^{mi}q^{nr})(q^{js}q^{pk} - q^{jk}q^{ps})(q^{lt}q^{qu} - q^{lu}q^{qt}) \\
&= (1/4!)\,(1/2)(-q)^{3/2}\, \underline{e}_{mnpq}\, \underline{e}_{rstu}\, q^{mr}q^{ni}\,(q^{js}q^{pk} - q^{jk}q^{ps})(q^{lt}q^{qu}) \\
&= (1/4!)\,(1/2)(-q)^{3/2}\, \Delta^{ijkl},
\end{aligned} \quad (A1)
$$

with

$$
\Delta^{ijkl} \equiv \underline{e}_{mnpq}\, \underline{e}_{rstu}\, q^{mr}q^{ni}\,(q^{js}q^{pk} - q^{jk}q^{ps})(q^{lt}q^{qu}), \quad (A2)
$$

we can write the Tamm-Rubilar tensor as

$$
G^{ijkl}(\chi) = Z^{(ijkl)}(\chi) = (1/4!)\,(1/2)\,(-q)^{3/2}\, \Delta^{(ijkl)}. \quad (A3)
$$

First calculate $\Delta^{1212}$ and use the equality $\hat{e}_{mnpq}\, q^{mr}q^{n1}\, q^{p1}\, q^{qu} = (\det q^{ij})\, \hat{e}^{r11u} = 0$ and

$\hat{e}_{mnpq} \hat{e}_{rstu} q^{mr} q^{n1} q^{ps} q^{qu} = (\det q^{ij}) \hat{e}^{r1su}$, we have:

$$\Delta^{1212} := \underline{e}_{mnpq} \underline{e}_{rstu} q^{mr} q^{n1} (q^{2s} q^{p1} - q^{21} q^{ps})(q^{2t} q^{qu}) = 6 (\det q^{ij}) q^{21} q^{21}. \tag{A4}$$

Similarly,

$$\Delta^{1112} = 6 (\det q^{ij}) q^{11} q^{21}; \quad \Delta^{1121} = 6 (\det q^{ij}) q^{11} q^{12};$$
$$\Delta^{1211} = 6 (\det q^{ij}) q^{11} q^{21}; \quad \Delta^{2111} = 6 (\det q^{ij}) q^{11} q^{12}. \tag{A5}$$

From these formulas, we have

$$\Delta^{(1112)} = (1/4)(\Delta^{1112} + \Delta^{1121} + \Delta^{1211} + \Delta^{2111}) = 6 (\det q^{ij}) q^{(11} q^{12)}. \tag{A6}$$

We note that in the total symmetrizing process $q^{11}$ is symmetric, i.e. $q^{11} = {}^{(S)}q^{11}$ and the antisymmetric parts of $q^{12}$ and $q^{21}$ cancel out, hence we have

$$q^{(11} q^{12)} = {}^{(S)}q^{(11 (S)} q^{12)}. \tag{A7}$$

Therefore, we have

$$\Delta^{(1112)} = 6 (\det q^{ij}) {}^{(S)}q^{(11 \; (S)} q^{12)}. \tag{A8}$$

From (A3), we have

$$G^{1112}(\chi) = (1/4!)(1/2)(-q)^{3/2} \Delta^{(1112)} = (1/8)(-q)^{3/2} (\det q^{ij}) ({}^{(S)}q^{(11(S)} q^{12)}). \tag{A9}$$

By calculation, we also have

$$G^{1111}(\chi) = (1/4!)(1/2)(-q)^{3/2} \Delta^{1111} = (1/8)(-q)^{3/2} (\det q^{ij}) ({}^{(S)}q^{11(S)} q^{12}). \tag{A10}$$

Similarly, we have

$$\Delta^{1122} = (\det q^{ij})(4 q^{21} q^{12} + 2 q^{11} q^{22});$$
$$\Delta^{1212} = 6 (\det q^{ij}) q^{21} q^{21};$$
$$\Delta^{1221} = (\det q^{ij})(4 q^{11} q^{22} + 2 q^{21} q^{12});$$
$$\Delta^{1122} + \Delta^{1212} + \Delta^{1221} = 6 (\det q^{ij})(q^{21} q^{21} + q^{21} q^{12} + q^{11} q^{22}). \tag{A11}$$

Applying permutation 1 ←→ 2 to obtain

$$\Delta^{2211} + \Delta^{2121} + \Delta^{2112} = 6 \ (\det q^{ij}) \ (q^{12}q^{12} + q^{12}q^{21} + q^{22}q^{11}), \tag{A12}$$

we have

$$\Delta^{(1122)} = 6 \ (\det q^{ij}) \ (q^{(11}q^{22)}) = 6 \ (\det q^{ij}) \ (^{(S)}q^{(11}q^{(S)22)}), \tag{A13}$$

and

$$G^{1122}(\chi) = (1/8)(-g)^{3/2} \det(q^{ij}) \ q^{(11}q^{22)} = (1/8) \ (-g)^{3/2} \det(q^{ij}) \ ^{(S)}q^{(11 \ (S)}q^{22)}. \tag{A14}$$

Similarly, we have by calculation

$$\begin{aligned}
\Delta^{1123} &= (\det q^{ij}) \ (4 \ q^{31}q^{12} + 2 \ q^{11}q^{32}); \\
\Delta^{1213} &= 6(\det q^{ij}) \ q^{31}q^{21}; \\
\Delta^{1231} &= (\det q^{ij}) \ (4 \ q^{11}q^{23} + 2 \ q^{21}q^{13}); \\
\Delta^{2113} &= (\det q^{ij}) \ (4 \ q^{32}q^{11} + 2 \ q^{12}q^{31}); \\
\Delta^{2131} &= 6(\det q^{ij}) \ q^{12}q^{13}; \\
\Delta^{2311} &= (\det q^{ij}) \ (4 \ q^{12}q^{31} + 2 \ q^{32}q^{11}) \text{ and permutations.}
\end{aligned} \tag{A15}$$

Therefore, we have

$$\Delta^{(1123)} = 6 \ (\det q^{ij}) \ (q^{(11}q^{23)}) = 6 \ (\det q^{ij}) \ (^{(S)}q^{(11}q^{(S)23)}), \tag{A16}$$

and

$$G^{1123}(\chi) = (1/8)(-q)^{3/2} \det(q^{ij}) \ q^{(11}q^{23)} = (1/8) \ (-q)^{3/2} \det(q^{ij}) \ ^{(S)}q^{(11 \ (S)}q^{23)}. \tag{A17}$$

Finally, we have

$$\Delta^{0123} = (\det q^{ij}) \ (4 \ q^{30}q^{12} + 2 \ q^{10}q^{32}), \tag{A18}$$

and

$$G^{0123} = (1/8) \ (-q)^{3/2} \ (\det q^{ij}) \ q^{(01}q^{23)} = 6(\det q^{ij}) \ ^{(S)}q^{(01}q^{(S)23)}, \tag{A19}$$

Putting (A9), (A10), (A14), (A17), (A19) and their permutations together, we obtain

$$G^{ijkl}(\chi) = (1/8) \ (-q)^{3/2} \det(q^{ij}) \ q^{(ij}q^{kl)} = (1/8) \ (-q)^{3/2} \det(q^{ij}) \ ^{(S)}q^{(ij \ (S)}q^{kl)}. \tag{A20}$$

**References**

[1] H. Minkowski, Konig. Ges. Wiss. Gottingen. Math.-Phys. (1908) 53-111.

[2] A. Einstein and M. Grossman, Zeit. Math. Phys. 62 (1913) 225-261.

[3] A. Einstein, Jarb. Radioakt. 4 (1907) 411-462.

[4] A. Einstein, Annalen der Physik 49 (1916) 769-822.

[5] A. Einstein, Eine Neue Formale Deutung der Maxwellschen Feldgleichungen der Elektrodynamik, *Königlich Preußische Akademie der Wissenschaften* (Berlin), 184-188 (1916); See also, A new formal interpretation of Maxwell's field equations of Electrodynamics, in *The Collected Papers of Albert Einstein*, Vol. 6, A. J. Kox et al., eds. (Princeton University Press: Princeton, 1996) pp. 263–269.

[6] See also F. W. Hehl, Ann. Phys. (Berlin) 17 (2008) 691-704 for a historical account and detailed explanation.

[7] H. Weyl, Raum-Zeit-Materie, Springer, Berlin, 1918; See also, Space-Time-Matter, English translation of the 4th German edition of 1922 (Dover, Mineola, New York, 1952).

[8] F. Murnaghan, The absolute significance of Maxwell's eruations, *Phys. Rev.* 17 (1921) 73-88.

[9] F. Kottler, Maxwell'sche Gleichungen und Metrik. *Sitzungsber. Akad. Wien IIa* 131 (1922) 119-146.

[10] É Cartan, Sur les variétés à connexion affine et la Théorie de la Relativité Généralisée, *Annales scientifirues de l'École Normale Supérieure* **40** pp. 325-412, **41** pp. 1-25, **42** pp. 17-88 (1923/1925); See also, On Manifolds with an Affine Connection and the Theory of General Relativity, English translation of the 1955 French edition. Bibliopolis, Napoli, 1986.

[11] F. W. Hehl and Yu. N. Obukhov, *Foundations of Classical Electrodynamics: Charge, Flux, and Metric* (Birkhäuser: Boston, MA, 2003).

[12] F. W. Hehl, Yu. N. Obukhov, G. F. Rubilar, On a possible new type of a T-odd skewon field linked to electromagnetism, in: A. Macias, F. Uribe, E. Diaz (Eds.), Developments in Mathematical and Experimental Physics, Volume A: Cosmology and Gravitation (Kluwer Academic/Plenum, New York, 2002) pp. 241-256 [gr-rc/0203096].

[13] R. Toupin, Elasticity and Electromagnetics, in Non-linear continuum theories, C.I.M.E. Conference, Bressanone, Italy (1965), C. Truesdell and G. Grioli, Eds., pp. 203-342.

[14] M. Schönberg, Electromagnetism and Gravitation, *Revista Brasileira de Fisica* **1** (1971) 91-122.

[15] A. Jadczyk, Electromagnetic permeability of the vacuum and light-cone structure,

  *Bulletin de l'Academie Polonaise des Sciences -- Séries des sciences physiques et astron.* 27 (1979) 91-94.

[16] W.-T. Ni, *Phys. Rev. Lett.* 38 (1977) 301.

[17] W.-T. Ni, *Bull. Am. Phys. Soc.* 19 (1974) 655.

[18] W.-T. Ni, Equivalence Principles, Their Empirical Foundations, and the Role of Precision Experiments to Test Them, in *Proceedings of the 1983 International School and Symposium on Precision Measurement and Gravity Experiment*, Taipei, Republic of China, January 24-February 2, 1983, ed. by W.-T. Ni (Published by National Tsing Hua University, Hsinchu, Taiwan, Republic of China, 1983) pp. 491-517.

[19] W.-T. Ni, Equivalence Principles and Precision Experiments, in *Precision Measurement and Fundamental Constants II*, ed. by B. N. Taylor and W. D. Phillips, Natl. Bur. Stand. (U S) Spec. Publ. 617 (1984) pp. 647-651 [http://astrod.wikispaces.com/].

[20] W.-T. Ni, Timing Observations of the Pulsar Propagations in the Galactic Gravitational Field as Precision Tests of the Einstein Equivalence Principle, in *Proceedings of the Second Asian-Pacific Regional Meeting of the International Astronomical Union on Astronomy, Bandung, Indonesia – 24 to29 August 1981*, ed. by B. Hidayat and M. W. Feast (Published by Tira Pustaka, Jakarta, Indonesia, 1984) pp. 441-448.

[21] M. Haugan and T. Kauffmann, Phys. Rev. D 52 (1995) 3168.

[22] C. Lämmerzahl and F. W. Hehl, *Phys. Rev. D* 70 (2004) 105022.

[23] W.-T. Ni, *Prog. Theor. Phys. Suppl.* **172** (2008) 49 [arXiv:0712.4082].

[24] W.-T. Ni, *Reports on Progress in Physics* 73 (2010) 056901.

[25] A. Favaro and L. Bergamin, Annalen der Physik 523 (2011) 383-401.

[26] M. F. Dahl, Journal of Physics A: Mathematical and Theoretical 45 (2012) 405203.

[27] D. Götz, S. Covino, A. Fernández-Soto, P. Laurent, and Ž. Bosnjak, Monthly Notice of Royal Astronomical Society 431 (2013) 3550.

[28] D. Götz et al., arXiv:1408.4121 (2014).

[29] W.-T. Ni, Equivalence principles, spacetime structure and the cosmic connection, to be published as Chapter 5 in the book: One Hundred Years of General Relativity: from Genesis and Empirical Foundations to Gravitational Waves, Cosmology and Quantum Gravity, edited by W.-T. Ni (World Scientific, Singapore, 2015).

[30] W.-T. Ni, Phys. Lett. A 378 (2014) 3413.

[31] S. di Serego Alighieri, W.-T. Ni, W.-P. Pan, Astrophys. J. 792 (2014) 35.

[32] S. di Serego Alighieri, Cosmological Birefringence: an Astrophysical test of

Fundamental Physics, Proceeding of Symposium I of JENAM 2010 – Joint European and National Astronomy Meeting: From Varying Couplings to Fundamental Physics, Lisbon, 6-10 Sept. 2010, Editors C. Martins and P. Molaro (Springer-Verlag, Berlin, 2011) p.139 [arXiv:1011.4865].

[33] W.-T. Ni, Implications of Hughes-Drever Experiments, in *Proceedings of the 1983 International School and Symposium on Precision Measurement and Gravity Experiment*, Taipei, Republic of China, January 24-February 2, 1983, ed. by W.-T. Ni (Published by National Tsing Hua University, Hsinchu, Taiwan, Republic of China, 1983) pp. 519-529 [http://astrod.wikispaces.com/].

[34] Yu. N. Obukhov and F. W. Hehl, Phys. Rev. D 70 (2004) 125015.

[35] W.-T. Ni, Skewon field and cosmic wave propagation, Phys. Lett. A 378 (2014) 1217-1223.

[36] A. S. Eddington, *The mathematical theory of relativity*, 2$^{nd}$ edition (Cambridge University Press, 1924).

[37] A. Einstein and E. G. Straus, Ann. Math. 47 (1946) 731.

[38] E. Schrödinger, Proc. R. Ir. Acad. 51A (1947) 163.

[39] E. Schrödinger, *Space-time structure* (Cambridge University Press, 1950).

[40] A. Favaro, Recent advances in classical electromagnetic theory, PhD thesis, Imperial College London, 2012.

[41] F. W. Hehl, Yu. N. Obukhov, G. F. Rubilar, Int. J. Mod. Phys. A 17 (2002) 2695.

[42] Yu. N. Obukhov and F. W. Hehl, Phys. Rev. D 70 (2004) 125015.

[43] F. W. Hehl and Yu. N. Obukhov, Phys. Lett. A 334 (2005) 249-259.

[44] F. W. Hehl, Yu. N. Obukhov, G. F. Rubilar and M. Blagojevic, Phys. Lett. A 347 (2005) 14.

[45] Y.Itin, J. Phys. A: Math. Theor. 42 (2009) 475402.

[46] Y. Itin, Phys. Rev. D 88 (2013) 107502.

[47] Y. Itin, Electromagnetic media with Higgs-type spontaneously broken transparency, arXiv:1406.3442v1.

[48] Y. Itin, On skewon modification of light cone structure, arXiv:1407.6722v1.

[49] W.-T. Ni, Foundations of Electromagnetism, Eruivalence Principles and Cosmic Interactions, Chaper 3 in *Trends in Electromagnetism - From Fundamentals to Applications*, pp. 45-68 (March, 2012), Victor Barsan and Radu P. Lungu (Ed.), ISBN: 978-953-51-0267-0, InTech (open access) (2012) [arXiv:1109.5501], Available from:
http://www.intechopen.com/books/trends-in-electromagnetism-from-fundamentals-to-applications/foundations-of-electromagnetism-eruivalence-principles-and-cosmic-interactions.

[50] W.-T. Ni, H.-H. Mei and S.-J. Wu, Mod. Phys. Lett. 28 (2013) 1340013.